\documentclass[fleqn,twoside]{article}
\usepackage{espcrc2}

\usepackage{graphicx}
\usepackage[figuresright]{rotating}

\newcommand{\AmS}{{\protect\the\textfont2
  A\kern-.1667em\lower.5ex\hbox{M}\kern-.125emS}}

\title{Topological flux sectors in extended U(1) gauge theory on $T^4$}

\author{Ph.\ de Forcrand%
  \address[ETH]{Institute for Theoretical Physics, 
    ETH Z\"{u}rich, CH-8093 Z\"{u}rich, Switzerland}%
  \address{CERN, Theory Division, CH-1211 Gen\`{e}ve 23, Switzerland}%
  \ and M. Vettorazzo%
  \addressmark[ETH] \thanks{Talk given by Michele Vettorazzo}%
  }

\begin{document}

\begin{abstract}
We consider the $4d$ compact $U(1)$ gauge theory with fundamental-adjoint action on a hypertorus. We give a full characterization of the
phase diagram of this model in terms of topological flux sectors.
\vspace{1pc}
\end{abstract}

\maketitle
\section{The flux in abelian gauge theories}
Consider an abelian gauge theory defined on a hypercube of size $L$ and periodic boundary
conditions.
In the continuum the definition of flux through any $(\mu,\nu)$ plane
is the following:
\begin{equation}
\Phi_{\mu\nu}=\oint F_{\mu\nu}d\sigma
\end{equation}
This quantity, because of periodic boundary conditions, is $2\pi
k$  valued ($k\in Z$), and configurations with different $k$ values are
topologically disconnected (so we
talk of topological superselection  $sectors$).

Consider now the same system, but with a discretized space-time.
The definition of flux is now
\begin{equation}
\Phi_{\mu\nu}=\frac{1}{L^2}\sum_{\mu\nu \hspace{0.1cm} planes}\sum_{P}\hspace{0.05cm}[\theta_{P_{\mu\nu}}]_{-\pi,\pi}
\end{equation}
where $[\theta_P]_{-\pi,\pi}$ is the plaquette angle reduced to the interval $(-\pi,\pi)$,
its so called $physical$ part.
A double sum is present: the internal $\sum_P$ is the sum over the plaquettes
in a single $(\mu,\nu)$ plane; this quantity is a multiple of $2\pi$, as in the continuum case.
But when we consider the external average over all parallel
planes, we observe that the flux can change from plane to plane,
due to the presence of $\emph{magnetic monopoles}$, specific to the lattice,
so that the allowed values for $\Phi_{\mu\nu}$ are multiples of $(2\pi)/L^2$.

Having in mind the continuum limit, configurations whose flux is $2\pi k$ play  a special role:
in fact $tunneling$ between different `sectors' is now possible, but higher and higher
barriers are expected between them as $L \to \infty$.
\begin{figure}[tb] 
\label{phase_diagram}
\centering
\includegraphics[height=5.0cm]{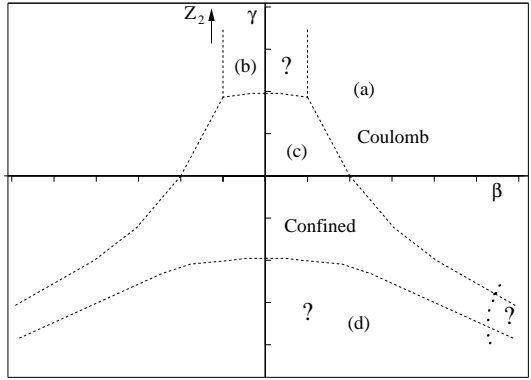}
\vspace*{-0.5cm}
\caption{Phase diagram of the extended $U(1)$ model.\vspace*{-0.25cm}}
\end{figure}
\vspace*{-0.25cm}
\section{The model}
\vspace*{-0.25cm}
We work with the following action
\begin{equation}
S=\beta \sum_P \cos \theta_{P} + \gamma \sum_P \cos 2\theta_{P}
\end{equation}
$\beta$ is the Wilson coupling. 
This model was  introduced by Bhanot \cite{Bhanot:1982zg} and has aroused interest also recently \cite{Campos:1998jp}.
In Fig.~1 we indicate three unknown aspects of the phase diagram with a question mark:\\
i) the nature of the phase indicated with $(b)$;\\
ii) the nature of the phase $(d)$; \\
iii) the fate of the two phase boundaries in the bottom right
corner of the phase diagram (the same holds on the left side, due to the symmetry $\beta\leftrightarrow -\beta$).
\vspace*{-0.25cm}
\section{Characterization of the phase transition}
\vspace*{-0.25cm}
Let us consider the behavior of flux sectors across a phase boundary;
here we work on the Wilson axis, but similar results hold everywhere.\\
The numerical strategy is very simple: using a $multicanonical$ approach, 
we measure the values of the flux through one $(\mu,\nu)$ orientation along
a Monte Carlo simulation and plot the (inverse) histogram of these 
measurements. The results are the following:\\ 
i) In the Coulomb phase flux sectors are well defined (Fig.~2),
and as the thermodynamic limit is approached we observe higher and higher barriers between them.\\
ii) In the confined phase (Fig.~3) the high density of monopoles hides 
the sectors and gives a Gaussian flux distribution. If we check
the thermodynamic limit, the distribution does $not$ change: in
fact more and more planes are present (this would give a narrower distribution), but also more and more
monopoles (which give a broader distribution) and these two effects compensate each other.
\begin{figure}[htb] 
\centering
\includegraphics[height=7.0cm,angle=-90]{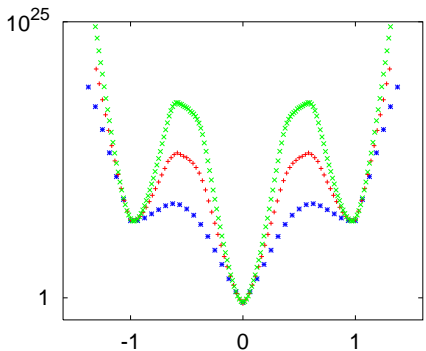}
\vspace*{-0.75cm}
\caption{Inverse distribution of the flux $\Phi_{\mu\nu}$ (eq. 2) in the Coulomb phase,  $\beta=1.2$ ($4^4,6^4,8^4$ lattice).\vspace*{-0.5cm}}
\end{figure}
\begin{figure}[htb] 
\centering
\includegraphics[height=6.5cm,angle=-90]{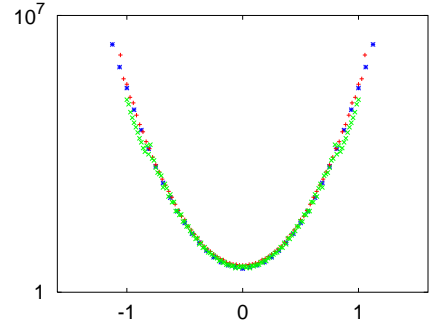}
\vspace*{-0.5cm}
\caption{Same as Fig.~2 in the confined phase, $\beta=0.8$.}
\vspace*{-0.5cm}
\end{figure}
\section{Characterization of the phases}
We extend to the abelian context the ideas of 't Hooft \cite{'tHooft:1979uj} about non-abelian gauge theories: we probe the
response of the system to a variation of the flux.
In the context of  $SU(N)$ pure gauge theories, we know that it is possible to modify the flux (by  twisting the boundary conditions) according to the discrete group $Z_N$.
We can implement a similar modification also in the abelian context, but a continuous variation of
the flux is now possible.

Our strategy is the following:
we consider a $stack$ of  plaquettes, one in each plane of a given orientation, and on these we change $\theta_P \to \theta_P+\phi$,
where $\phi\in [0,2\pi]$. This corresponds to imposing an extra flux $\phi$ through the chosen orientation.
The new partition function is
\begin{equation}
Z(\phi)=\int D\theta\hspace{0.05cm} e^{\beta (\sum_{stack}\cos (\theta_p+\phi)+
\sum_{\overline{stack}}\cos \theta_p)}
\end{equation}
and analogously away from the Wilson axis; $Z(\phi)$ is clearly $2\pi$ periodic. 
We measure the free energy of the flux, $F(\phi)=-\log \frac{Z(\phi)}{Z(0)}$.

\begin{figure}
\centering
\includegraphics[height=6.0cm,angle=-90]{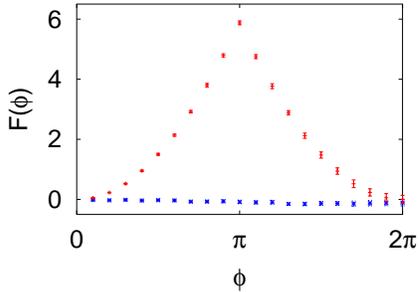}
\vspace*{-0.5cm}
\caption{Free energy $F(\phi)$ in phase $(a)$ (upper curve) and phase $(c)$ 
(lower curve) .}
\end{figure}
\begin{figure}[htb]
\vspace*{-0.5cm} 
\centering
\includegraphics[height=6.5cm,angle=-90]{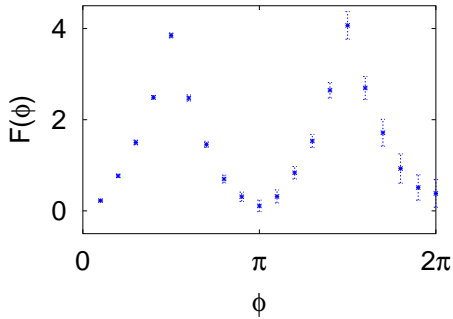}
\vspace{-0.5cm}
\caption{Typical free energy profile in phases $(b)$ and $(d)$.}
\end{figure}
We make the following observations: \\
i) In the Coulomb phase we recover the $2\pi$ sectors (Fig.~4, upper curve).\\
ii) In the confined phase (Fig.~4, lower curve), within our statistical error, the free energy is independent of $\phi$. This is due to the decoupling of flux values through different planes, so that the effects of the variation of the flux are screened.\\
iii) In the unknown phases $(b),(d)$ we observe the appearance of an extra $\pi$ periodicity (Fig.~5). We present an argument to interpret this result:
consider the following quantity
\begin{equation}
\langle P_q(x)P_q^\dagger(x+L\hat e_\mu)\rangle_{\tilde Z}
\end{equation}
where $P_q$ is the Polyakov loop of charge $q$, $\tilde Z=\int_0^{2\pi}d\phi Z(\phi)$, and $\hat e_\mu$ is the unit vector in direction $\hat\mu$.
This correlator is non trivial due to 'twisted' b.c..
Gauss law ($P_q(x+L\hat e_\mu)= e^{-iq\phi}P_q(x)$) gives
\begin{equation}
\langle P_q(x)P_q^\dagger(x+L\hat e_\mu)\rangle_{\tilde Z}=
\frac{\int_0^{2\pi}d\phi e^{iq\phi}  Z(\phi)}{\int_0^{2\pi}d\phi Z(\phi)}
\end{equation}
If $Z(\phi)$ has a $\pi$ periodicity, it follows that when $q=1$, or more generally odd, the correlator is zero, while if $q$ is even, it can be different
 from zero. We then remember that $\langle P_qP_q^\dagger\rangle\sim e^{V_{eff}(L)T}$,
so we deduce that odd charges are confined (in even combinations). We therefore claim that these two unknown phases $(b),(d)$ are just Coulomb phases for even charges.
\vspace*{-0.35cm}
\section{Fate of the phase boundaries}
\vspace*{-0.2cm}
The last question we address regards the fate
of the two phase boundaries at the lower corners of Fig.~1. It is possible that they meet at some
point or that they just become asymptotically close to each other.
What we can exclude is that they end somewhere, letting the two
different Coulomb phases communicate.

Suppose that they communicate somewhere, then we expect the  
twisted free energy $F(\phi)$  to be the same in both: this leads to a contradiction. 
In fact $F(\phi)$ 
would have $least$ periodicity $\pi$ and $2\pi$ at the same time. This is possible
if the distribution is flat; but this again implies confinement.
\vspace*{-0.35cm}
\section{Conclusions and perspectives}
\vspace*{-0.2cm}
We observe that in the literature about $U(1)$ the issue
of ergodicity through flux sectors is usually not considered, despite the early observation of \cite{Grosch:1985cz}; in the region close to criticality this phenomenon
could be relevant.

Possible perspectives of this work are the following: the signal
related to the different periodicities of the partition function is
very strong; it is clearly visible on small lattices also at large
$\beta$. We can use it  as a tool to study the position of
the phase boundaries, and so to extend the quantitative picture of
the phase diagram.\\ Secondly, we want to try an FSS analysis of
the quantity $F(\pi)$ (the free energy with twist-flux $\Phi=\pi$)
as we cross a phase boundary: indications about the nature of the
phase transition ($1^{st}$ or $2^{nd}$ order) could
be obtained.\\

\vspace*{-0.2cm}
We gratefully acknowledge Oliver Jahn for useful discussions.

\end{document}